\documentclass[12pt]{article}
\usepackage[margin=1in]{geometry}
\usepackage{times}
\usepackage{color}
\usepackage{graphicx}
\usepackage[colorlinks,citecolor=blue,urlcolor=blue,linkcolor=blue,bookmarks=false,hypertexnames=true]{hyperref} 

\usepackage{multirow}
\usepackage{amsmath}
\usepackage{amsthm}
\usepackage{amssymb}
\usepackage{natbib}
\usepackage{rotating}
\usepackage{appendix}
\usepackage{titling}  

\usepackage{authblk}

\DeclareMathOperator*{\argmin}{arg\,min}

\newtheorem{prop}{Proposition}
\newtheorem{lem}{Lemma}

\begin{document}
\title{A Note on Harmonic Underspecification in Log-Normal Trigonometric Regression}
\author[1]{Michael T. Gorczyca\footnote{Corresponding author: mtg62@cornell.edu}}

\affil[1]{MTG Research Consulting}

\date{\today}
\maketitle

\begin{abstract}
\noindent Analysis of biological rhythm data often involves performing least squares trigonometric regression, which models the oscillations of a response over time as a sum of sinusoidal components. When the response is not normally distributed, an investigator will either transform the response before applying least squares trigonometric regression or extend trigonometric regression to a generalized linear model (GLM) framework. In this note, we compare these two approaches when the number of oscillation harmonics is underspecified. We assume data are sampled under an equispaced experimental design and that a log link function would be appropriate for a GLM. We show that when the response follows a generalized gamma distribution, least squares trigonometric regression with a log-transformed response, or log-normal trigonometric regression, produces unbiased parameter estimates for the oscillation harmonics, even when the number of oscillation harmonics is underspecified. In contrast, GLMs require correct specification to produce unbiased parameter estimates. We apply both methods to cortisol level data and find that only log-normal trigonometric regression produces parameter estimates that are invariant to the number of specified oscillation harmonics. Additionally, when a sufficiently large number of oscillation harmonics is specified, both methods produce identical parameter estimates for the oscillation harmonics.
\end{abstract}

\section{Introduction}

Many biological processes, such as feeding \citep{Pickel2020}, sleep-wake cycles \citep{Lavie2001}, and menstrual cycles \citep{Oguz2013}, occur at predictable time points relative to their oscillation periods. These biological rhythms have attracted increasing research attention over the past two decades, with studies identifying novel associations between biomarker oscillations and human health \citep{Hughes2017}. In analyzing biological rhythm data, a scenario that has become common involves regression with non-normally distributed responses. Biological rhythm researchers have addressed this either by transforming the response and applying least squares regression \citep{Dolu2019, Gupta2022, Zong2023} or by fitting a generalized linear model (GLM) with an appropriate distribution and link function \citep{Doyle2022, Parsons2024, Velikajne2022}.

In this note, we consider these two approaches in the context of trigonometric regression, which models the oscillations of a response as a sum of sinusoidal components \citep{Bingham1982, Cornelissen2014, Tong1976}. We focus on a scenario in which the number of oscillation harmonics is underspecified, or fewer oscillation harmonics are included in the model than are necessary to fully represent the oscillations of a response. This underspecification is done to produce a parsimonious model and interpretable hypothesis test results \citep{Madden2018}. We first review least squares trigonometric regression, GLMs, and response transformation in Section \ref{sec:2}. We then illustrate each approach using cortisol level data in Section \ref{sec:3}. Finally, we conclude with a brief discussion in Section \ref{sec:4}.

\section{Background} \label{sec:2}

\subsection{Least Squares Trigonometric Regression} \label{sec:2.1}

Suppose an investigator conducts a biological rhythm experiment, collecting $n$ independent samples $\{(X_i, Y_i), i=1,\ldots,n\}$ of the random variables $(X, Y)$, where $Y_i$ denotes a measurement of a phenomenon (the response) at time $X_i$. After sample collection, the investigator will typically specify a trigonometric regression model of order $K$ to represent the response over time, or assume
\begin{align}
Y_i = \mu_0 + \sum_{k=1}^K\mu_k\cos(k\omega X_i + \phi_k) + \varepsilon_i. \label{eq:amp_phase}
\end{align}
In (\ref{eq:amp_phase}), $\mu_0$ represents the midline of oscillation for the response, the quantity $2\pi/\omega$ represents a known oscillation period, and each term $\mu_k\cos(k\omega X + \phi_k)$ for $k \in \{1,\ldots,K\}$ represents an oscillation harmonic. For the $k$th oscillation harmonic, $\mu_k$ denotes an amplitude parameter, or the deviation from the midline to the $k$th oscillation harmonic's peak; and $\phi_k$ denotes a phase-shift parameter, which relates to the times at which the $k$th oscillation harmonic peaks relative to the oscillation period \citep{Cornelissen2014, Gorczyca2026}. The term $\varepsilon_{i}$ in (\ref{eq:amp_phase}) represents random noise associated with the $i$th measurement that is independent of $X_i$. The expectation of this random noise $\mathbb{E}(\varepsilon_{i}) = 0$ and the variance $\mathrm{Var}(\varepsilon_{i}) = \sigma^2$. It is noted that the true order parameter for the amplitude-phase model in (\ref{eq:amp_phase}) will be defined as $K^*$. We claim the model in (\ref{eq:amp_phase}) is underspecified when $K< K^*$, and that the model is correctly specified when $K\geq K^*$.

While the amplitude-phase model in (\ref{eq:amp_phase}) is biologically interpretable, the model is nonlinear with respect to its amplitudes and phase-shifts, which would complicate parameter estimation without prior knowledge about the true parameters \citep[Theorem 6.7]{Boos2013}. To avoid making assumptions about the true parameters, many investigators instead specify the linear model
\begin{align}
    Y_i =  f(X_i)^T\beta + \varepsilon_i = \beta_0 + \sum_{k=1}^K \left\{ \beta_{2k-1} \sin\left( k \omega X_i\right) + \beta_{2k} \cos\left(k \omega  X_i\right) \right\} + \varepsilon_i, \label{eq:lin_mod}
\end{align}
where
\begin{align*}
\beta = \begin{bmatrix}\beta_0 &  \beta_1 & \dots & \beta_{2K} \end{bmatrix}^T
\end{align*}
is a vector of unknown parameters and
\begin{align*}
f(X_i) = \begin{bmatrix}1 & \sin\left(\omega X_i\right) & \cos\left(\omega X_i\right) & \dots & \sin\left(K \omega X_i\right) & \cos\left(K \omega X_i\right)\end{bmatrix}^T
\end{align*}
is a vector of regression functions \citep{Cornelissen2014, Gorczyca2025, Tong1976, Bingham1982}. The parameters of the linear model in (\ref{eq:lin_mod}) can be transformed to the parameters of the amplitude-phase model in (\ref{eq:amp_phase}). Specifically, $\mu_0$ in (\ref{eq:amp_phase}) is equal to $\beta_0$ in (\ref{eq:lin_mod}), and for each $k$th oscillation harmonic of the model in (\ref{eq:amp_phase}),
\begin{equation*} \begin{aligned}
    \mu_k = \sqrt{\beta_{2k-1}^2+\beta_{2k}^2}, \ \quad & \phi_k = \mathrm{atan2}(-\beta_{2k-1}, \beta_{2k}),  \\
    \beta_{2k-1} = -\mu_k\sin(\phi_k), \quad \quad \quad  & \beta_{2k} = \mu_k\cos(\phi_k),
\end{aligned} \label{eq:alt_to_orig} \end{equation*}
where $\mathrm{atan2}(\cdot, \cdot)$ denotes the two-argument arctangent function. 

To estimate the parameters of the linear model in (\ref{eq:lin_mod}), an investigator will apply least squares, or solve the optimization problem
\begin{align}
\hat{\beta} = \argmin_{B \in \mathbb{R}^{2K+1}} \frac{1}{n} \sum_{i=1}^n \left\{ Y_i - f(X_i)^TB \right\}^2. \label{eq:ls}
\end{align}
If the investigator assumes
\begin{align}
    Y_i \sim \mathcal{N}\{f(X_i)^T\beta, \sigma^2\}, \label{eq:Yn}
\end{align}
then the covariance matrix of $\hat{\beta}$ obtained in (\ref{eq:ls}) is defined as
\begin{align}
\mathrm{Cov}(\hat{\beta}) &= \hat{\sigma}^2 \left\{ \sum_{i=1}^n f(X_i) f(X_i)^T \right\}^{-1}, \label{eq:cov}
\end{align}
where 
\begin{align*}
    \hat{\sigma}^2 &= \frac{1}{n-2K-1} \sum_{i=1}^n \left\{ Y_i - \hat{\beta}^T f(X_i) \right\}^2
\end{align*}
denotes an estimate of the residual variance, or the variance of the random noise if the model is correctly specified \citep[Section 6.1]{Moser1996}.

Previous research has investigated how different sample collection protocols affect the precision of parameter estimates and the power of hypothesis tests for the linear model in (\ref{eq:lin_mod}) \citep[Pages 241-243]{Pukelsheim2006}. In this note, we introduce two assumptions based on this research:
\begin{description}
    \item[Assumption 1.] Samples are collected at equally spaced time points relative to the oscillation period of the response, or under an equispaced experimental design, with $X_i =  2\pi(i-1)/(\omega n)$ for all $i \in \{1,\ldots, n\}$.
    \item[Assumption 2.] The sample size $n>2K^*$, where $K^*$ denotes the true order parameter of the model in (\ref{eq:amp_phase}).
\end{description}
Assumption 1 is that samples are collected under an equispaced experimental design, which is the optimal experimental design for maximizing the precision in parameter estimation and power in hypothesis testing for trigonometric regression under multiple statistical criteria \citep[Pages 241-243]{Pukelsheim2006}. Equispaced experimental designs have also been recommended for biological rhythm studies \citep{Hughes2017}. Notably, Assumption 1 implies the covariance matrix in (\ref{eq:cov}) would simplify to the diagonal matrix
\begin{align*}
\mathrm{Cov}(\hat{\beta}) = \hat{\sigma}^2
\begin{bmatrix}
\frac{1}{n} & 0 & \cdots & 0 \\
0 & \frac{1}{2n} & \cdots & 0 \\
0 & 0 & \ddots & 0 \\
0 & 0 & \cdots & \frac{1}{2n}
\end{bmatrix}.
\end{align*}
As a consequence, each parameter estimate is mutually independent (\citealp{Cornelissen2014}; \citealp{Gorczyca2025}; \citealp[Lemma 1.7]{Tsybakov2009}). Assumption 2 would ensure that all the true oscillation harmonics fall below the Nyquist limit and avoids aliasing, or that the parameter estimates are identifiable when the model is correctly specified \citep[Pages 21-23]{Bloomfield2000}. Both of these assumptions imply that if the model in (\ref{eq:lin_mod}) is underspecified with $K<K^*$, then $\hat{\beta}$ is unbiased under expectation. 

\subsection{Generalized Linear Models} \label{sec:2.2}

In some biological rhythm studies, the assumption that the response $Y_i$ follows the normal distribution in (\ref{eq:Yn}) could be inappropriate. In this scenario, investigators have applied trigonometric regression within a generalized linear model (GLM) framework \citep{Doyle2022, Parsons2024, Velikajne2022}. As a brief overview, define $\eta_i = f(X_i)^T\beta$ as the linear predictor for the $i$th sample and let $\mu_i = \mathbb{E}(Y_i \mid X_i)$ denote the corresponding conditional mean. In a GLM, the quantities $\eta_i$ and $\mu_i$ are related through a link function $g(\cdot)$ such that
\begin{align}
\mu_i = g^{-1}(\eta_i). \label{eq:glm_mean}
\end{align}
GLMs require the assumption that the conditional distribution of $Y_i$ given $X_i$ belongs to the exponential dispersion model (EDM) family. Distributions in this family have probability density or mass functions of the form
\begin{align}
P(Y_i;\theta_i,\phi)
= \exp\left\{\frac{Y_i \theta_i - b(\theta_i)}{\phi} + c(Y_i,\phi)\right\}, \label{eq:EDM}
\end{align}
where $\theta_i$ is a quantity referred to as the canonical parameter, $\phi>0$ is a dispersion parameter, $b(\cdot)$ is a cumulant function, and $c(\cdot,\cdot)$ is a normalizing function. The conditional mean and variance of $Y_i$ given $X_i$ can be expressed as
\begin{align*}
\mu_i = \frac{d b(\theta_i)}{d \theta_i}, &\quad
\mathrm{Var}(Y_i \mid X_i) = \phi \frac{d^2 b(\theta_i)}{d \theta_i^2}.
\end{align*}
It is noted that if the link function $g(\cdot)$ also satisfies the relationship
\begin{align*}
    g(\mu_i) = \theta_i,
\end{align*}
then $g(\cdot)$ is referred to as the canonical link function for a distribution \citep[Chapter 5]{Dunn2018}.

In general, parameter estimation for a GLM does not admit a closed-form solution. However, the covariance matrix of $\hat{\beta}$ for a GLM is defined as
\begin{align}
\mathrm{Cov}(\hat{\beta})
= \left\{ \sum_{i=1}^{n} w_i f(X_i) f(X_i)^{T} \right\}^{-1}, \label{eq:cov2}
\end{align}
where the $i$th weight
\begin{align*}
w_i
= \frac{(d \mu_i/d \eta_i)^2}{\mathrm{Var}(Y_i \mid X_i)},
\end{align*}
which depends on both the variance of $Y_i$ and the link function specified \citep[Page 126]{Agresti2015}. Unlike the scenario where $Y_i$ follows a normal distribution and an identity link function is specified, these weights generally vary across samples. As a result, even if Assumptions 1 and 2 are valid, the regression functions $f(X_i)$ would need to be mutually orthogonal with respect to the weights. If the regression functions are not mutually orthogonal with respect to these weights, then an underspecified GLM-based trigonometric regression can produce biased parameter estimates.

\subsection{Transformation of the Response} \label{sec:2.3}

An alternative to GLMs for handling non-normally distributed responses in biological rhythm data is to transform the response prior to least squares \citep{Dolu2019, Gupta2022, Zong2023}. The predominant approach for transforming the response for biological rhythm data is to apply a log transform, in which an investigator would instead estimate the parameters of the model in (\ref{eq:lin_mod}) by computing 
\begin{align*}
\hat{\beta} = \argmin_{B \in \mathbb{R}^{2K+1}} \frac{1}{n} \sum_{i=1}^n \left\{ \log(Y_i) - B^T f(X_i) \right\}^2,
\end{align*}
where $\log(\cdot)$ denotes the natural logarithm \citep{Dolu2019, Gupta2022}. It is noted that this transform corresponds to the assumption that $Y_i$ follows a log-normal distribution, and will refer to regression with this transformed response as log-normal trigonometric regression.

One benefit of transforming the response and then performing least squares trigonometric regression is that, under Assumptions 1 and 2, the parameter estimates of a correctly specified model would be mutually orthogonal. As a result, if the investigator assumes the model is correctly specified with respect to a transform, which would be assumed for inference \citep{Box1982}, then the parameter estimates would be unbiased under expectation when Assumptions 1 and 2 are valid, even if the investigator underspecifies $K<K^*$. 

We identify an additional benefit to performing log-normal trigonometric regression instead of fitting a GLM. Specifically, some non-normal distributions the response could follow include the chi-squared, gamma, and Weibull distributions. Each of these distributions, as well as the log-normal distribution, can be expressed as special cases of the generalized gamma distribution \citep{Stacy1962}, which we define as having a conditional probability density function of the form
\begin{align}
    P(Y_i \mid \lambda(X_i), \kappa, \rho)=\frac{\rho Y_i^{\kappa-1} \exp\left[-\left\{Y_i/\lambda(X_i)\right\}^p\right]}{\lambda(X_i)^\kappa\Gamma\left(\kappa/\rho\right)}, \label{eq:gg}
\end{align}
where
\begin{align}
    \lambda(X_i) &= \frac{\Gamma\left(\kappa/\rho\right)\exp\left\{f^*(X_i)^T\beta^* \right\}}{\Gamma\left\{(\kappa+1)/\rho\right\}} \label{eq:lamb}
\end{align}
and
\begin{align*}
    \Gamma(u)=\int_0^{\infty}v^{u-1}\exp(-v)dv.
\end{align*}
It is noted that in (\ref{eq:lamb}),
\begin{align*}
f^*(X_i) = \begin{bmatrix}1 & \sin\left(\omega X_i\right) & \cos\left(\omega X_i\right) & \dots & \sin\left(K^* \omega X_i\right) & \cos\left(K^* \omega X_i\right)\end{bmatrix}^T
\end{align*}
is a correctly specified vector of regression functions, and $\beta^*$ is the corresponding $(2K^*+1)\times 1$ vector of unknown parameters. The following proposition shows that log-normal trigonometric regression can produce unbiased parameter estimates for the oscillation harmonics when the response follows a generalized gamma distribution.
\begin{prop} \label{prop:1}
Suppose the response $Y_i$ follows the generalized gamma distribution defined in (\ref{eq:gg}). If an investigator computes the parameters of a log-normal trigonometric regression model of order $K \leq K^*$ when both Assumptions 1 and 2 are valid, then
\begin{align*}
    \mathbb{E}(\hat{\beta}) &= \begin{bmatrix} (\beta^*_0  +c_0) & \beta^*_{1} & \ldots & \beta^*_{2K-1} & \beta^*_{2K}\end{bmatrix}^T,
\end{align*}
where the constant
\begin{align*}
    c_0 = \frac{\psi\left(\kappa/\rho\right)}{\rho}
- \log\left[\Gamma\left\{(\kappa+1)/\rho\right\}\right] + \log\left\{\Gamma\left(\kappa/\rho\right)\right\}
\end{align*}
with 
\begin{align*}
    \psi(u) = \frac{d \log\{\Gamma(u)\}}{d u}.
\end{align*}
\end{prop}
\noindent The derivation of this result is provided in Appendix \ref{app:A}. The implication of Proposition \ref{prop:1} is that if an investigator assumes $Y_i$ follows a generalized gamma distribution and specifies a log link function in a GLM framework, then the investigator would obtain biased parameter estimates when the number of oscillation harmonics is underspecified. Under these same assumptions, the log-normal trigonometric regression model would produce unbiased parameter estimates for the oscillation harmonics under expectation, even if the number of oscillation harmonics is underspecified. It is noted that hypothesis testing in biological rhythm studies focuses on the amplitudes and the phase shifts, for which unbiased estimates can be obtained from the log-normal model using the identities in (\ref{eq:alt_to_orig}) \citep{Bingham1982, Tong1976}.

\section{Illustration with Cortisol Level Data} \label{sec:3}

We analyze cortisol level data from an experiment in which blood samples were collected every two hours over a 24-hour period from three cohorts: nine healthy individuals with no known illnesses, eleven individuals diagnosed with major depressive disorder, and sixteen individuals with Cushing’s syndrome \citep{Wang1996}. For illustration, we focus on the cohort of healthy individuals and fit second-order ($K=2$) and fifth-order ($K=5$) models separately for each individual (for each individual-level model, the sample size $n=12$). We consider $K=2$, which has been identified in a separate analysis of this cortisol level data \citep{Albert2005}, and $K=5$, which is the largest number of oscillation harmonics that can be specified under Assumptions 1 and 2. For modeling the response, we compare a gamma GLM with a log link function to a least squares model where the response was log-transformed. It is noted that one of these nine individuals had a missing sample, and this individual's data was removed from the analysis.

Table \ref{tab:1} presents the parameter estimates obtained for a log-normal model and a gamma GLM. For the log-normal model, the parameter estimates for the first two oscillation harmonics are invariant to whether a second-order or fifth-order model is fit. The gamma GLM, on the other hand, produces different parameter estimates depending on the order term specified. Table \ref{tab:1} also shows that both fifth-order models produce the same parameter estimates for the oscillation harmonics. It is noted that these results are consistent with Proposition \ref{prop:1}.

\section{Discussion} \label{sec:4}

We examined the effect of underspecifying the number of oscillation harmonics on parameter estimation for trigonometric regression with a non-normally distributed response. We found that, when samples are collected under an equispaced experimental design, GLM parameter estimates are generally not mutually orthogonal, unlike those obtained from least squares. As a consequence, a GLM can produce biased parameter estimates when the number of oscillation harmonics is underspecified. To address this limitation, we consider transforming the response and then performing least squares. Proposition \ref{prop:1} supports this approach, showing that log-normal trigonometric regression can produce unbiased parameter estimates for the oscillation harmonics under expectation, even when the model is underspecified. Further, our cortisol level data illustration in Section \ref{sec:3} numerically shows that the parameter estimates for a log-normal model are invariant to the number of oscillation harmonics specified.

\newpage 
\clearpage 
\begin{table*}[!h]
	\caption{Parameter estimates for individual-level models. ``LT'' denotes least squares trigonometric regression with a log-transformed response, and ``GLM'' denotes trigonometric regression extended to a gamma generalized linear model. LT produces identical parameter estimates for the first two oscillation harmonics ($\hat{\beta}_k$ for $k=1,\ldots, 4$), regardless of the order term specified ($K=2$ or $K=5$). GLM, on the other hand, produces order-dependent parameter estimates. Both fifth-order models (LT and GLM) produce identical parameter estimates for the oscillation harmonics ($\hat{\beta}_k$ for $k=1,\ldots, 10$).} \label{tab:1}
 \centering
 \resizebox{1.05\textwidth}{!}{
\begin{tabular}{|c| cc|cc|cc|cc|cc|cc|cc|cc|}\hline 
\multicolumn{17}{|c|}{Second-Order Model $(K=2)$} \\
\hline
Subject ID & \multicolumn{2}{c|}{8001} & \multicolumn{2}{c|}{8002} & \multicolumn{2}{c|}{8003} & \multicolumn{2}{c|}{8004} & \multicolumn{2}{c|}{8005} & \multicolumn{2}{c|}{8006} & \multicolumn{2}{c|}{8008} & \multicolumn{2}{c|}{8009} \\
\hline 
Model & LT & GLM & LT & GLM & LT & GLM & LT & GLM & LT & GLM & LT & GLM & LT & GLM & LT & GLM \\
\hline 
$\hat{\beta}_0$& 4.966 & 5.207 & 3.590 & 4.015 & 3.662 & 3.776 & 3.539 & 3.913 & 3.626 & 3.719 & 2.804 & 3.038 & 3.660 & 3.798 & 3.556 & 3.953 \\ 
$\hat{\beta}_1$& 1.442 & 1.082 & 1.529 & 1.656 & 1.370 & 1.347 & 2.080 & 1.942 & 1.059 & 1.105 & 2.149 & 2.164 & 2.610 & 2.559 & 1.144 & 0.791 \\  
$\hat{\beta}_2$& 0.875 &  0.919 &  1.323 &  1.543 &  1.140 &  1.086 &  0.337 &  0.319 &  2.104 &  2.142 &  1.134 &  1.342 & -0.239 & -0.153 &  0.034 & -0.184 \\ 
$\hat{\beta}_3$& -0.102 & -0.182 &  0.275 &  0.145 &  0.327 &  0.403 &  0.567 &  0.537 & -1.375 & -1.364 & -1.216 & -1.280 &  1.516 &  1.404 &  0.691 &  0.915 \\  
$\hat{\beta}_4$& 1.000 &  0.813 &  1.576 &  1.945 &  0.968 &  1.035 &  1.092 &  1.143 &  0.949 &  0.965 &  1.246 &  1.310 &  0.648 &  0.705 & -0.742 & -0.685 \\ \hline
\multicolumn{17}{|c|}{Fifth-Order Model $(K=5)$} \\
\hline
Subject ID & \multicolumn{2}{c|}{8001} & \multicolumn{2}{c|}{8002} & \multicolumn{2}{c|}{8003} & \multicolumn{2}{c|}{8004} & \multicolumn{2}{c|}{8005} & \multicolumn{2}{c|}{8006} & \multicolumn{2}{c|}{8008} & \multicolumn{2}{c|}{8009} \\
\hline 
Model & LT & GLM & LT & GLM & LT & GLM & LT & GLM & LT & GLM & LT & GLM & LT & GLM & LT & GLM \\
\hline 
$\hat{\beta}_0$&4.966 & 5.026 & 3.590 & 3.681 & 3.662 & 3.662 & 3.539 & 3.542 & 3.626 & 3.628 & 2.804 & 2.845 & 3.660 & 3.687 & 3.556 & 3.556 \\ 
$\hat{\beta}_1$&1.442 & 1.442 & 1.529 & 1.529 & 1.370 & 1.370 & 2.080 & 2.080 & 1.059 & 1.059 & 2.149 & 2.149 & 2.610 & 2.610 & 1.144 & 1.144 \\ 
$\hat{\beta}_2$&0.875 &  0.875 &  1.323 &  1.323 &  1.140 &  1.140 &  0.337 &  0.337 &  2.104 &  2.104 &  1.134 &  1.134 & -0.239 & -0.239 &  0.034 &  0.034 \\ 
$\hat{\beta}_3$&-0.102 & -0.102 &  0.275 &  0.275 &  0.327 &  0.327 &  0.567 &  0.567 & -1.375 & -1.375 & -1.216 & -1.216 &  1.516 &  1.516 &  0.691 &  0.691 \\ 
$\hat{\beta}_4$&1.000 &  1.000 &  1.576 &  1.576 &  0.968 &  0.968 &  1.092 &  1.092 &  0.949 &  0.949 &  1.246 &  1.246 &  0.648 &  0.648 & -0.742 & -0.742 \\ 
$\hat{\beta}_5$&0.315 &  0.315 &  0.466 &  0.466 &  0.308 &  0.308 &  0.270 &  0.270 &  0.212 &  0.212 & -0.182 & -0.182 &  0.343 &  0.343 &  0.576 &  0.576 \\ 
$\hat{\beta}_6$&-0.004 & -0.004 & -0.634 & -0.634 & -0.128 & -0.128 &  0.175 &  0.175 & -0.130 & -0.130 & -0.548 & -0.548 &  0.271 &  0.271 & -0.590 & -0.590 \\ 
$\hat{\beta}_7$&-0.223 & -0.223 & -0.763 & -0.763 &  0.207 &  0.207 &  0.409 &  0.409 &  0.237 &  0.237 & -0.018 & -0.018 &  0.250 &  0.250 &  0.726 &  0.726 \\ 
$\hat{\beta}_8$&0.436 &  0.436 & -0.670 & -0.670 &  0.254 &  0.254 &  1.100 &  1.100 & -0.517 & -0.517 & -0.473 & -0.473 &  0.426 &  0.426 &  0.695 &  0.695 \\ 
$\hat{\beta}_9$&-0.576 & -0.576 &  0.109 &  0.109 & -0.037 & -0.037 & -0.048 & -0.048 & -0.112 & -0.112 &  0.639 &  0.639 & -0.147 & -0.147 & -0.222 & -0.222 \\ 
$\hat{\beta}_{10}$&-0.401 & -0.401 & -0.551 & -0.551 & -0.466 & -0.466 & -0.226 & -0.226 &  0.014 &  0.014 & -0.311 & -0.311 &  0.008 &  0.008 & -0.242 & -0.242 \\
\hline
\end{tabular}
}
\end{table*}

\newpage 
\clearpage 

\appendix

\section{Theory Results} \label{app:A}

\subsection{Supporting Lemma}

\begin{lem}[Equation 17, \citealt{Ashkar1998}] \label{lem:1}
Let $Y_i$ be a realization of the random variable $Y$. If the probability density function of $Y_i$ follows a generalized gamma distribution, or
\begin{align*}
    P(Y_i \mid \lambda, \kappa, \rho)=\frac{\rho Y_i^{\kappa-1} \exp\left\{-\left(Y_i/\lambda\right)^\rho \right\}}{\lambda^\kappa\Gamma\left(\kappa/\rho\right)}, \label{eq:gg}
\end{align*}
then 
\begin{align*}
\mathbb{E}\left\{\log(Y)\right\} &= \log(\lambda) + \frac{\psi\left(\kappa/\rho\right)}{\rho}.
\end{align*}
\end{lem}

\subsection{Proposition \ref{prop:1} Derivation}
We are given
\begin{align*}
\mu_i = \mathbb{E}(Y_i \mid X_i) = \exp\{f^*(X_i)^T \beta^*\},
\end{align*}
and that the conditional distribution of $Y_i$ given $X_i$ follows a generalized gamma distribution. Lemma \ref{lem:1} coupled with the definition of the generalized gamma distribution in (\ref{eq:gg}) implies that
\begin{align*}
\mathbb{E}\{\log(Y_i) \mid X_i\}
&= \log\{\lambda(X_i)\} + \frac{\psi\left(\kappa/\rho\right)}{\rho} \\
&= \log\left[\frac{\Gamma\left(\kappa/\rho\right)\exp\left\{f^*(X_i)^T\beta^* \right\}}{\Gamma\left((\kappa+1)/\rho\right)}\right] + \frac{\psi\left(\kappa/\rho\right)}{\rho}\\
&= f^*(X_i)^T\beta^* + \log\left\{\Gamma\left(\kappa/\rho\right)\right\} - \log\left[\Gamma\left\{(\kappa+1)/\rho\right\}\right] + \frac{\psi\left(\kappa/\rho\right)}{\rho} \\
&= f^*(X_i)^T \beta^* + c_0,
\end{align*}
where
\begin{align*}
    c_0 &= \log\left\{\Gamma\left(\frac{\kappa}{\rho}\right)\right\} - \log\left[\Gamma\left\{\frac{(\kappa+1)}{\rho}\right\}\right] + \frac{\psi\left(\kappa/\rho\right)}{\rho}.
\end{align*}
Now, define the parameter vector
\begin{align*}
\tilde{\beta}
&= \begin{bmatrix}
\beta^*_0 + c_0 & \beta^*_1 & \ldots & \beta^*_{2K^*}
\end{bmatrix},
\end{align*}
and recall that the estimate $\hat{\beta}$ obtained by least squares in (\ref{eq:ls}) can also be identified by solving the normal equations
\begin{align*}
\left\{\sum_{i=1}^n f(X_i) f(X_i)^T \right\} \hat{\beta}
-
\left\{\sum_{i=1}^n f(X_i)\log(Y_i)\right\}
= 0.
\end{align*}
If we also define the $(2K+1)\times(2K+1)$ matrix
\begin{align*}
W_{2K+1}
&= \sum_{i=1}^n f(X_i) f(X_i)^T
= \mathrm{diag}_{2K+1}\!\left(n,\frac{n}{2},\ldots,\frac{n}{2}\right),
\end{align*}
where the diagonal structure follows from orthogonality of distinct terms in a trigonometric basis from Assumptions 1 and 2 \citep[Lemma 1.7]{Tsybakov2009}, then taking the expectation of the normal equations yields
\begin{align*}
\mathbb{E}\left[
\left\{\sum_{i=1}^n f(X_i) f(X_i)^T\right\}\hat{\beta}
-
\left\{\sum_{i=1}^n f(X_i)\log(Y_i)\right\}
\right]
&=
W_{2K+1}\mathbb{E}(\hat{\beta})
-
\sum_{i=1}^n f(X_i)\mathbb{E}\{\log(Y_i)\mid X_i\} \\
&=
W_{2K+1}\mathbb{E}(\hat{\beta})
-
\sum_{i=1}^n f(X_i)\{f^*(X_i)^T\tilde{\beta}\}.
\end{align*}
We can simplify the term
\begin{align*}
\sum_{i=1}^n f(X_i)\{f^*(X_i)^T\tilde{\beta}\}
&=
\begin{bmatrix}
n(\beta^*_0 + c_0) &
\frac{n\beta^*_1}{2} &
\frac{n\beta^*_2}{2} &
\ldots &
\frac{n\beta^*_{2K-1}}{2} &
\frac{n\beta^*_{2K}}{2}
\end{bmatrix},
\end{align*}
again by orthogonality of the trigonometric basis, and conclude
\begin{align*}
\mathbb{E}(\hat{\beta})
=
\begin{bmatrix}
(\beta^*_0 + c_0) &
\beta^*_1 &
\ldots &
\beta^*_{2K-1} &
\beta^*_{2K}
\end{bmatrix}.
\end{align*}

\bibliographystyle{apalike} 
\bibliography{bibliography}

@BOOK{Agresti2015,
  title     = "Foundations of linear and generalized linear models",
  author    = "Agresti, Alan",
  publisher = "John Wiley \& Sons",
  series    = "Wiley Series in Probability and Statistics",
  month     =  feb,
  year      =  2015,
  address   = "Nashville, TN",
  language  = "en"
}

@article{Albert2005,
  title = {On Analyzing Circadian Rhythms Data Using Nonlinear Mixed Models with Harmonic Terms},
  volume = {61},
  ISSN = {1541-0420},
  url = {http://dx.doi.org/10.1111/j.0006-341X.2005.464_1.x},
  DOI = {10.1111/j.0006-341x.2005.464_1.x},
  number = {4},
  journal = {Biometrics},
  publisher = {Oxford University Press (OUP)},
  author = {Albert,  Paul S. and Hunsberger,  Sally},
  year = {2005},
  month = dec,
  pages = {1115–1120}
}

@article{Ashkar1998,
  title = {Approximate Confidence Intervals for Quantiles of Gamma and Generalized Gamma Distributions},
  volume = {3},
  ISSN = {1943-5584},
  url = {http://dx.doi.org/10.1061/(ASCE)1084-0699(1998)3:1(43)},
  DOI = {10.1061/(asce)1084-0699(1998)3:1(43)},
  number = {1},
  journal = {Journal of Hydrologic Engineering},
  publisher = {American Society of Civil Engineers (ASCE)},
  author = {Ashkar,  Fahim and Ouarda,  Taha B. M. J.},
  year = {1998},
  month = jan,
  pages = {43–51}
}

@article{Bingham1982,
    title = {Inferential statistical methods for estimating and comparing cosinor parameters},
    author = {Bingham, C. and Arbogast, B. and Cornelissen-Guillaume, G. G. and Lee, J. K. and Halberg, F.},
    year = {1982},
    journal = {Chronobiologia},
    pages = {397-439},
    volume = {9},
    issue = {4}
}

@book{Bloomfield2000,
  title = {Fourier Analysis of Time Series: An Introduction},
  ISBN = {9780471722236},
  ISSN = {1940-6347},
  url = {http://dx.doi.org/10.1002/0471722235},
  DOI = {10.1002/0471722235},
  journal = {Wiley Series in Probability and Statistics},
  publisher = {Wiley},
  author = {Bloomfield,  Peter},
  year = {2000},
  month = jan 
}

@BOOK{Boos2013,
  title     = "Essential statistical inference",
  author    = "Dennis D. Boos and Leonard A. Stefanski",
  publisher = "Springer",
  series    = "Springer Texts in Statistics",
  edition   =  2013,
  month     =  feb,
  year      =  2013,
  address   = "New York, NY",
  language  = "en"
}

@article{Box1982,
  title = {An Analysis of Transformations Revisited,  Rebutted},
  volume = {77},
  ISSN = {1537-274X},
  url = {http://dx.doi.org/10.1080/01621459.1982.10477788},
  DOI = {10.1080/01621459.1982.10477788},
  number = {377},
  journal = {Journal of the American Statistical Association},
  publisher = {Informa UK Limited},
  author = {Box,  G. E. P. and Cox,  D. R.},
  year = {1982},
  month = mar,
  pages = {209–210}
}

@article{Cornelissen2014,
  doi = {10.1186/1742-4682-11-16},
  year = {2014},
  month = apr,
  publisher = {Springer Science and Business Media {LLC}},
  volume = {11},
  number = {1},
  author = {Germaine Cornelissen},
  title = {Cosinor-based rhythmometry},
  journal = {Theoretical Biology and Medical Modelling}
}

@article{Dolu2019,
  title = {Feasibility of the duration of actigraphy data to illustrate circadian rhythm among cognitively intact older people in nursing home: cosinor analysis},
  volume = {18},
  ISSN = {1479-8425},
  url = {http://dx.doi.org/10.1007/s41105-019-00245-w},
  DOI = {10.1007/s41105-019-00245-w},
  number = {1},
  journal = {Sleep and Biological Rhythms},
  publisher = {Springer Science and Business Media LLC},
  author = {Dolu,  Ilknur and Nahcivan,  Nursen O.},
  year = {2019},
  month = nov,
  pages = {59–64}
}

@article{Doyle2022,
  title = {Enhancing cosinor analysis of circadian phase markers using the gamma distribution},
  volume = {92},
  ISSN = {1389-9457},
  url = {http://dx.doi.org/10.1016/j.sleep.2022.01.015},
  DOI = {10.1016/j.sleep.2022.01.015},
  journal = {Sleep Medicine},
  publisher = {Elsevier BV},
  author = {Doyle,  Margaret M. and Murphy,  Terrence E. and Miner,  Brienne and Pisani,  Margaret A. and Lusczek,  Elizabeth R. and Knauert,  Melissa P.},
  year = {2022},
  month = apr,
  pages = {1–3}
}

@book{Dunn2018,
  title = {Generalized Linear Models With Examples in R},
  ISBN = {9781441901187},
  ISSN = {2197-4136},
  url = {http://dx.doi.org/10.1007/978-1-4419-0118-7},
  DOI = {10.1007/978-1-4419-0118-7},
  journal = {Springer Texts in Statistics},
  publisher = {Springer New York},
  author = {Dunn,  Peter K. and Smyth,  Gordon K.},
  year = {2018}
}

@article{Gorczyca2025,
  doi = {10.1002/sim.70201},
  year = {2025},
  author = {Gorczyca, Michael T. and Sefas, Justice D.},
  title = {Weighted Trigonometric Regression for Suboptimal Designs in Circadian Transcriptome Studies},
  journal = {Statistics in Medicine},
  publisher = {Wiley},
  volume = {44},
  number = {20-22}
}

@article{Gorczyca2026,
  title = {A mixed-effects cosinor modelling framework for circadian gene expression},
  volume = {620},
  ISSN = {0022-5193},
  url = {http://dx.doi.org/10.1016/j.jtbi.2025.112301},
  DOI = {10.1016/j.jtbi.2025.112301},
  journal = {Journal of Theoretical Biology},
  publisher = {Elsevier BV},
  author = {Gorczyca,  Michael T.},
  year = {2026},
  month = mar,
  pages = {112301}
}

@article{Gupta2022,
  title = {Circadian rest‐activity misalignment in critically ill medical intensive care unit patients},
  volume = {31},
  ISSN = {1365-2869},
  url = {http://dx.doi.org/10.1111/jsr.13587},
  DOI = {10.1111/jsr.13587},
  number = {5},
  journal = {Journal of Sleep Research},
  publisher = {Wiley},
  author = {Gupta,  Prerna and Martin,  Jennifer L. and Malhotra,  Atul and Bergstrom,  Jaclyn and Grandner,  Michael A. and Kamdar,  Biren B.},
  year = {2022},
  month = apr 
}

@article{Hughes2017,
  doi = {10.1177/0748730417728663},
  year = {2017},
  month = oct,
  publisher = {{SAGE} Publications},
  volume = {32},
  number = {5},
  pages = {380--393},
  author = {Michael E. Hughes and Katherine C. Abruzzi and Ravi Allada and Ron Anafi and Alaaddin Bulak Arpat and Gad Asher and Pierre Baldi and Charissa de Bekker and Deborah Bell-Pedersen and Justin Blau and Steve Brown and M. Fernanda Ceriani and Zheng Chen and Joanna C. Chiu and Juergen Cox and Alexander M. Crowell and Jason P. DeBruyne and Derk-Jan Dijk and Luciano DiTacchio and Francis J. Doyle and Giles E. Duffield and Jay C. Dunlap and Kristin Eckel-Mahan and Karyn A. Esser and Garret A. FitzGerald and Daniel B. Forger and Lauren J. Francey and Ying-Hui Fu and Fr{\'{e}}d{\'{e}}ric Gachon and David Gatfield and Paul de Goede and Susan S. Golden and Carla Green and John Harer and Stacey Harmer and Jeff Haspel and Michael H. Hastings and Hanspeter Herzel and Erik D. Herzog and Christy Hoffmann and Christian Hong and Jacob J. Hughey and Jennifer M. Hurley and Horacio O. de la Iglesia and Carl Johnson and Steve A. Kay and Nobuya Koike and Karl Kornacker and Achim Kramer and Katja Lamia and Tanya Leise and Scott A. Lewis and Jiajia Li and Xiaodong Li and Andrew C. Liu and Jennifer J. Loros and Tami A. Martino and Jerome S. Menet and Martha Merrow and Andrew J. Millar and Todd Mockler and Felix Naef and Emi Nagoshi and Michael N. Nitabach and Maria Olmedo and Dmitri A. Nusinow and Louis J. Pt{\'{a}}{\v{c}}ek and David Rand and Akhilesh B. Reddy and Maria S. Robles and Till Roenneberg and Michael Rosbash and Marc D. Ruben and Samuel S.C. Rund and Aziz Sancar and Paolo Sassone-Corsi and Amita Sehgal and Scott Sherrill-Mix and Debra J. Skene and Kai-Florian Storch and Joseph S. Takahashi and Hiroki R. Ueda and Han Wang and Charles Weitz and P{\aa}l O. Westermark and Herman Wijnen and Ying Xu and Gang Wu and Seung-Hee Yoo and Michael Young and Eric Erquan Zhang and Tomasz Zielinski and John B. Hogenesch},
  title = {Guidelines for genome-scale analysis of biological rhythms},
  journal = {Journal of Biological Rhythms},
}

@article{Lavie2001,
  title = {Sleep-Wake as a Biological Rhythm},
  volume = {52},
  ISSN = {1545-2085},
  url = {http://dx.doi.org/10.1146/annurev.psych.52.1.277},
  DOI = {10.1146/annurev.psych.52.1.277},
  number = {1},
  journal = {Annual Review of Psychology},
  publisher = {Annual Reviews},
  author = {Lavie,  P.},
  year = {2001},
  month = feb,
  pages = {277–303}
}

@article{Madden2018,
  title = {Morning surge in blood pressure using a random‐effects multiple‐component cosinor model},
  volume = {37},
  ISSN = {1097-0258},
  url = {http://dx.doi.org/10.1002/sim.7607},
  DOI = {10.1002/sim.7607},
  number = {10},
  journal = {Statistics in Medicine},
  publisher = {Wiley},
  author = {Madden,  J.M. and Browne,  L.D. and Li,  X. and Kearney,  P.M. and Fitzgerald,  A.P.},
  year = {2018},
  month = jan,
  pages = {1682–1695}
}

@book{Moser1996,
    author = {Moser, Barry K.},
    title = {Linear models: a mean model approach},
    publisher = {Academic Press},
    year = 1996
}

@article{Oguz2013,
  title = {Effects of menstrual cycle showing infradian rhythm on thyroid blood flow and thyroid volume in healthy women},
  volume = {44},
  ISSN = {1744-4179},
  url = {http://dx.doi.org/10.1080/09291016.2011.652863},
  DOI = {10.1080/09291016.2011.652863},
  number = {1},
  journal = {Biological Rhythm Research},
  publisher = {Informa UK Limited},
  author = {Oguz,  Ayten and Gumus,  Mehmet and Ipek,  Ali and Tuzun,  Dilek and Ersoy,  Reyhan and Cakir,  Bekir},
  year = {2013},
  month = feb,
  pages = {103–112}
}

@article{Parsons2024,
  title = {GLMMcosinor: Flexible Cosinor Modeling to Characterize Rhythmic Time Series Using a Generalized Linear Mixed Modeling Framework},
  url = {http://dx.doi.org/10.1101/2024.04.10.588934},
  DOI = {10.1101/2024.04.10.588934},
  publisher = {Cold Spring Harbor Laboratory},
  author = {Parsons,  Rex and Jayasinghe,  Oliver and White,  Nicole and Chunduri,  Prasad and Rawashdeh,  Oliver},
  year = {2024},
  journal = {bioRxiv},
  month = apr 
}

@article{Pickel2020,
  title = {Feeding Rhythms and the Circadian Regulation of Metabolism},
  volume = {7},
  ISSN = {2296-861X},
  url = {http://dx.doi.org/10.3389/fnut.2020.00039},
  DOI = {10.3389/fnut.2020.00039},
  journal = {Frontiers in Nutrition},
  publisher = {Frontiers Media SA},
  author = {Pickel,  Lauren and Sung,  Hoon-Ki},
  year = {2020},
  month = apr 
}

@BOOK{Pukelsheim2006,
  title     = "Optimal Design of Experiments",
  author    = "Friedrich Pukelsheim",
  publisher = "Society for Industrial and Applied Mathematics",
  series    = "Springer Texts in Statistics",
  month     =  jan,
  year      =  2006,
  address   = "Philadelphia, PA"
}

@article{Stacy1962,
  title = {A Generalization of the Gamma Distribution},
  volume = {33},
  ISSN = {0003-4851},
  url = {http://dx.doi.org/10.1214/aoms/1177704481},
  DOI = {10.1214/aoms/1177704481},
  number = {3},
  journal = {The Annals of Mathematical Statistics},
  publisher = {Institute of Mathematical Statistics},
  author = {Stacy,  E. W.},
  year = {1962},
  month = sep,
  pages = {1187–1192}
}

@article{Tong1976,
  title = {Parameter Estimation in Studying Circadian Rhythms},
  volume = {32},
  ISSN = {0006-341X},
  url = {http://dx.doi.org/10.2307/2529340},
  DOI = {10.2307/2529340},
  number = {1},
  journal = {Biometrics},
  publisher = {JSTOR},
  author = {Tong,  Y. L.},
  year = {1976},
  month = mar,
  pages = {85}
}

@book{Tsybakov2009,
  title = {Introduction to Nonparametric Estimation},
  ISBN = {9780387790527},
  ISSN = {0172-7397},
  url = {http://dx.doi.org/10.1007/b13794},
  DOI = {10.1007/b13794},
  journal = {Springer Series in Statistics},
  publisher = {Springer New York},
  author = {Tsybakov,  Alexandre B.},
  year = {2009}
}

@article{Velikajne2022,
  title = {RhythmCount: A Python package to analyse the rhythmicity in count data},
  volume = {63},
  ISSN = {1877-7503},
  url = {http://dx.doi.org/10.1016/j.jocs.2022.101758},
  DOI = {10.1016/j.jocs.2022.101758},
  journal = {Journal of Computational Science},
  publisher = {Elsevier BV},
  author = {Velikajne,  Nina and Moškon,  Miha},
  year = {2022},
  month = sep,
  pages = {101758}
}

@article{Wang1996,
  title = {A Flexible Model for Human Circadian Rhythms},
  volume = {52},
  ISSN = {0006-341X},
  url = {http://dx.doi.org/10.2307/2532897},
  DOI = {10.2307/2532897},
  number = {2},
  journal = {Biometrics},
  publisher = {JSTOR},
  author = {Wang,  Yuedong and Brown,  Morton B.},
  year = {1996},
  month = jun,
  pages = {588-596}
}

@article{Zong2023,
  doi = {10.1002/sim.9803},
  year = {2023},
  month = jun,
  publisher = {Wiley},
  volume = {42},
  number = {18},
  pages = {3236--3258},
  author = {Wei Zong and Marianne L. Seney and Kyle D. Ketchesin and Michael T. Gorczyca and Andrew C. Liu and Karyn A. Esser and George C. Tseng and Colleen A. McClung and Zhiguang Huo},
  title = {Experimental design and power calculation in omics circadian rhythmicity detection using the cosinor model},
  journal = {Statistics in Medicine}
}

\end{document}